
\magnification1200

\font\BBBig=cmr10 scaled\magstep3
\font\small=cmr8

\font\twelve=cmbx10


\def\title{
{\bf\BBBig
\centerline{Conformal Properties}\medskip \centerline{of}\medskip
\centerline{Chern-Simons Vortices}\medskip \centerline{in}\medskip
\centerline{External Fields}
}}

\def\foot#1{
\footnote{($^{\the\foo}$)}{#1}\advance\foo by 1 } 
\def\ccr{\cr\noalign{\medskip}}


\def\authors{
\centerline{
C. DUVAL\foot{\small and D\'epartement de Physique,
Universit\'e d'Aix-Marseille II,
UFR de Luminy, Case 907, F--13288 MARSEILLE Cedex 09
(France),
e-mail~: duval@cpt.univ-mrs.fr.}
P. A. HORV\'ATHY\foot{\small
Permanent address~: D\'epartement de Math\'ematiques,
Universit\'e de Tours, Parc de Grandmont, 
F--37200 TOURS (France), e-mail~: horvathy@univ-tours.fr} L. PALLA\foot{\small
Permanent address~: Institute for Theoretical Physics,
E\"otv\"os University, H--1088 BUDAPEST, 
Puskin u. 5-7 (Hungary),
e-mail~: palla@ludens.elte.hu}}}

\def\runningauthors{Duval, Horv\'athy, Palla}
\def\runningtitle{Time-dependent Chern-Simons\dots}


\voffset = 1cm 
\baselineskip = 14pt 

\headline ={
\ifnum\pageno=1\hfill
\else\ifodd\pageno\hfil\tenit\runningtitle\hfil\tenrm\folio
\else\tenrm\folio\hfil\tenit\runningauthors\hfil \fi
\fi}

\nopagenumbers
\footline={\hfil} 


\def\and{\qquad\hbox{and}\qquad}

\def\kikezd{\parag\underbar}

\def\IR{{\bf R}}

\def\smallcirc{{\raise 0.5pt \hbox{$\scriptstyle\circ$}}}
\def\smallover#1/#2{\hbox{$\textstyle{#1\over#2}$}}
\def\2{{\smallover1/2}}
\def\ccr{\cr\noalign{\medskip}}
\def\parag{\hfil\break}
\def\semidirectproduct{
{\ooalign{\hfil\raise.07ex\hbox{s}\hfil\crcr\mathhexbox20D}} }
\def\={\!=\!}
\def\boxit#1{
\vbox{\hrule\hbox{\vrule\kern3pt
\vbox{\kern3pt#1\kern3pt}\kern3pt\vrule}\hrule} } 

\def\cJ{{\cal J}}
\def\cB{{\cal G}}
\def\cG{{\cal G}}
\def\cP{{\cal P}}
\def\cD{{\cal D}}
\def\cK{{\cal K}}
\def\cH{{\cal H}}
\def\cM{{\cal N}}


\newcount\ch 
\newcount\eq 
\newcount\foo 
\newcount\ref 

\def\chapter#1{
\parag\eq = 1\advance\ch by 1{\bf\the\ch.\enskip#1} }

\def\equation{
\leqno(\the\ch.\the\eq)\global\advance\eq by 1 }

\def\reference{
\parag [\number\ref]\ \advance\ref by 1
}

\ch = 0 
\foo = 1 
\ref = 1 


\centerline{\twelve CENTRE DE PHYSIQUE THEORIQUE - CNRS - Luminy,
Case 907}
\centerline{\twelve F--13288 Marseille Cedex 9 - France }
\centerline{ Unit\'e Propre de Recherche 7061}
\vskip 1.5cm

\title
\vskip 1.5cm
\authors
\vskip .25in

\parag
\centerline{{\bf Abstract.}}

{\it The construction and the symmetries of Chern-Simons vortices
in harmonic and uniform magnetic force backgrounds found by Ezawa,
Hotta and Iwazaki, and by Jackiw and Pi are generalized using the
non-relativistic Kaluza-Klein-type framework presented in our
previous paper. All Schr\"odinger-symmetric backgrounds are determined.}

\vskip.2in


\vskip.2in


\noindent
April 1994

\noindent
CPT-94/P.3028

\medskip

\noindent
anonymous ftp or gopher~: cpt.univ-mrs.fr

\bigskip

\vfill\eject


\chapter{Introduction}

The construction of static, non-relativistic Chern-Simons solitons
[1] was recently generalized to time-dependent solutions, yielding
vortices in a constant external magnetic field, ${\cal B}$ [2-4].
 Putting $\omega={\cal B}/2$, the equation to be solved is
$$
i\big({D_\omega}\big)_t\Psi_\omega=\left\{ -{1\over2}{\vec{D}}_\omega^2
-\Lambda\,\Psi_\omega^*\Psi_\omega
\right\}\Psi_\omega.
\equation
$$
(We use units where $e=m=1$). Here the covariant derivative means
$$
\big({D_\omega}\big)_\alpha
\=
\partial_\alpha-i({A_\omega})_\alpha-i{\cal A}_\alpha \equation
$$
($\alpha=0,1,2$), where ${\cal A}_\alpha$ is a vector potential
for the constant magnetic field, ${\cal A}_0=0$, ${\cal A}_i=
\2\epsilon_{ij}x^j{\cal B}\equiv\omega\epsilon_{ij}x^j$ ($i,j=1,2$)
and
$(A_\omega)_\alpha$ is the vector potential of Chern-Simons
electromagnetism i.e. its field strength is required to satisfy
 the field-current identity
$$
B_\omega\equiv
\epsilon^{ij}\partial_iA_\omega^j
=-{1\over\kappa}\rho_\omega
\and
E_\omega^i\equiv
-\partial_iA_\omega^0-\partial_tA_\omega^i ={1\over\kappa}
\epsilon^{ij}J_\omega^j
\equation
$$
with
$\rho_\omega\=\Psi_\omega^*\Psi_\omega$
and
$\vec{J}_\omega\=
({1/2i})[\Psi^*\vec{D}_\omega
\Psi_\omega-\Psi_\omega(\vec{D}_\omega\Psi_\omega)^*] $.
These equations can be solved [2-4] by applying a coordinate
transformation to a solution, $\Psi$, of the problem with
$\omega=0$ studied in Ref. [1], according to
$$\eqalign{
\Psi_\omega(t,\vec{x})&={1\over\cos\omega t}\,
\exp\left\{-i\omega{r^2\over2}\tan{\omega t}\right\}\,
\exp\left\{i{{\cal N}\over2\pi\kappa}\omega t\right\}\,
\Psi(\vec{X},T),
\ccr
(A_\omega)_\alpha&=A_\beta{\partial X^\beta\over\partial
x^\alpha} -\partial_\alpha\Big({\omega\over2\pi\kappa}
{\cal N}t\Big), \cr}
\equation
$$
with
$$
T={\tan\omega t\over\omega},
\qquad
\vec{X}={1\over\cos\omega t}R(\omega t)\,\vec{x}. \equation
$$
Here ${\cal N}=\int\Psi^*\Psi\,d^2\vec{x}$ is the vortex number
 and $R(\theta)$ is the matrix of a rotation by angle $\theta$
 in the plane.
(The pre-factor $\exp[i{\cal N}\omega t/2\pi\kappa]$ and the
extra term $-\partial_\alpha[(\omega/2\pi\kappa){\cal N}t]$
are absent from the corresponding formula of Ezawa et al. [2]).
 A similar construction works in a harmonic background [4].

In a previous paper [5]
non-relativistic Chern-Simons theory
in $2+1$ dimensions was obtained by reduction from an
appropriate $(3+1)$-dimensional Lorentz manifold. As
an application, we reproduced the results in Ref. [1].
Here we show that the above generalizations arise by
reduction from suitable curved spaces, that they all
share the (extended) Schr\"odinger symmetry of the model
in [1], and we determine all background fields which have
this property.

\goodbreak


\chapter{Chern-Simons theory in Bargmann space}

$(2+1)$-dimensional non-relativistic Chern-Simons theory
 can be lifted to \lq Bargmann space' i.e. to a $4$-dimensional
 Lorentz manifold $(M,g)$ endowed with a
covariantly constant null vector $\xi$ [5]. The theory is
described by a massless non-linear wave equation
$$
\Big\{D_\mu D^\mu
-{R\over6}
+\lambda\psi^*\psi\Big\}\psi=0,
\equation
$$
where $D_\mu=\nabla_\mu-ia_\mu$ ($\mu=0,1,2,3$), $\nabla$
 is the metric-covariant derivative and $R$ denotes the
scalar curvature.
The scalar field $\psi$ and the
`electromagnetic' field strength,
$f_{\mu\nu}=2\partial_{[\mu}a_{\nu]}$,
are related by the identity
$$
\kappa f_{\mu\nu}=
\sqrt{-g}\,\epsilon_{\mu\nu\rho\sigma}\xi^\rho j^\sigma,
\qquad
j^\mu={1\over2i}\left[\psi^*(D^\mu\psi)-\psi(D^\mu\psi)^*\right].
 \equation
$$
Eq.~(2.1) [but {\it not} (2.2)] can be obtained from variation
of the `partial' action
$$
S={1\over2}\int_M\left\{(D_\mu\psi)^*\,D^\mu\psi
+{R\over 6}\,|\psi|^2-{\lambda\over 2}\,|\psi|^4 \right\}
\sqrt{-g}\,d^4\!x.
\equation
$$

The quotient of $M$ by the integral curves of $\xi$ is
non-relativistic space-time we denote by $Q$. A Bargmann
space admits local coordinates $(t,\vec{x},s)$ such that
$(t,\vec{x})$ label $Q$ and
$\xi=\partial_s$.
When supplemented by the equivariance condition
$$
\xi^\mu D_\mu\psi=i\psi,
\equation
$$
our theory projects to a non-relativistic non-linear
Schr\"odinger/Chern-Simons
theory on the $(2+1)$-dimensional manifold $Q$. The
field strength $f_{\mu\nu}$ is clearly the lift of a closed
two-form $F_{\mu\nu}$ on $Q$. So, the vector potential may be
chosen as $a_\mu=(A_\alpha,0)$ with $A_\alpha$ $s$-independent.
 In this gauge, $\Psi(t,\vec{x})=e^{-is}\psi(t,\vec{x})$ is then
 a function on $Q$.

A {\it symmetry} is a transformation of $M$ which interchanges
the solutions of the coupled system. Each $\xi$-preserving
 conformal transformation is a symmetry and the variational
 derivative
$\vartheta_{\mu\nu}=2\delta S/\delta g^{\mu\nu}$ provides
us with a conserved,
traceless and symmetric energy-momentum tensor. The version of
Noether's theorem proved in [5] says that for any
$\xi$-preserving conformal vectorfield
$(X^\mu)$ on Bargmann space, the quantity
$$
Q_X=\int_{\Sigma_t}{
\vartheta_{\mu\nu}X^\mu\xi^\nu\,\sqrt{\gamma}\,d^2\vec{x} }
\equation
$$
(where the `transverse space' $\Sigma_t$ is a space-like
2-surface $t={\rm const.}$ and
$\gamma_{ij}$ is the metric induced on it by $g_{\mu\nu}$)
is a constant of the motion. The conserved quantities are
conveniently calculated using the formula [5]
$$
\vartheta_{\mu\nu}\xi^\nu
={1\over2i}\left[
\psi^*\,(D_\mu\psi)-\psi\,(D_\mu\psi)^*\right]
-{1\over 6}\,\xi_\mu\left(
{R\over 6}|\psi|^2
+(D^\nu\psi)^*\,D_\nu\psi
+{\lambda\over2}\,|\psi|^4
\right).
\equation
$$

For example, $M$ can be flat Minkowski space with metric
$d\vec{X}^2+2dTdS$, where
$\vec{X}\in\IR^2$ and $S$ and $T$ are light-cone coordinates.
 This is the
Bargmann space of a free, non-relativistic particle [6]. The
system of equations (2.1,2,4) projects in this case to that of
 Ref. [1]; the
$\xi$-preserving conformal transformations form the (extended)
 planar Schr\"o\-din\-ger group, consisting of the Galilei
group with generators $\cJ$ (rotation),
$\cH$ (time translation),
$\vec{\cG}$ (boosts),
$\vec{\cP}$ (space translations),
augmented with dilatation, $\cD$, and expansion, $\cK$, and
centrally extended by `vertical' translation, $\cM$. With a
slight abuse of notation, the associated conserved quantities
 are denoted by the same symbols.
(Explicit formul{\ae} are listed in [1] and [5]). Applying
any symmetry transformation to a solution of the field equations
 yields another solution. For example, a boost or an expansion
applied to the static solution $\Psi_0(\vec{X})$ of Jackiw and
Pi produces time-dependent solutions. Using the formul{\ae}
in [6] we find
$$
\Psi(T,\vec{X})={1\over1-kT}\exp\left\{-{i\over2}\Big[
2\vec{X}\cdot\vec{b}+T\vec{b}^2+k{(\vec{X}+\vec{b}T)^2\over1-kT}
 \Big]\right\}\,\Psi_0({\vec{X}+\vec{b}T\over1-kT}), \equation
$$
which is the same as in [1].

Now we present some new results.
The most general `Bargmann' manifold was found long ago by
Brinkmann [7]:
$$
g_{ij}dx^idx^j+2dt\big[ds+\vec{\cal A}\cdot d\vec{x}\big]
+2{\cal A}_0dt^2,
\qquad
{\cal A}_0=-U
\equation
$$
where the `transverse' metric $g_{ij}$ as well as the
`vector potential' $\vec{\cal A}$ and the `scalar' potential
 $U$ are functions of $t$ and $\vec{x}$ only. Clearly,
$\xi=\partial_s$ is a covariantly constant null vector.
The null geodesics of this metric describe particle motion
 in curved transverse space in an external electromagnetic
fields $\vec{\cal E}=-\partial_t\vec{\cal A}-\vec{\nabla}U$
 and ${\cal B}=\vec{\nabla}\times\vec{\cal A}$ [6].

Consider now a Chern-Simons vector potential
$(a_\omega)_\mu=\big((A_\omega)_\alpha,0\big)$
in the background (2.8).
[The subscript $(\,\cdot\,)_\omega$ refers to an
external-field problem].
Using that the only non-vanishing components of the metric
 (2.8) are $g^{ij},\, g^{is}=-A^i$,
$g^{ss}=2U+A_iA^i$,
$g^{st}=1$,
we find that the integrand in the partial action (2.3) is
$$
\big(\vec{D}_\omega\psi\big)^*\cdot\vec{D}_\omega\psi
+i\Big[\big((D_\omega)_t\psi\big)^*\psi
-\psi^*(D_\omega)_t\psi\Big]
+{R\over6}\,|\psi|^2-{\lambda\over 2}\,|\psi|^4, \equation
$$
where the covariant derivative $(D_\omega)_\alpha$ means
(1.2) with vector
potential ${\cal A}_\alpha$. Thus, including the `vector-potential'
 components into the metric (2.8) results, after reduction,
simply in modifying the covariant derivative $D_\alpha$ in `empty'
 space (${\cal A}_\alpha=0$) according to
$
D_\alpha\to(D_\omega)_\alpha.
$
The associated equation of motion is hence Eq.~(1.1). (This
conclusion can also be reached
directly by studying the wave equation (2.1)).

Let now $\varphi$ denote a conformal Bargmann diffeomorphism
between two
Bargmann spaces i.e. let $\varphi:(M,g,\xi)\to(M',g',\xi')$
be such that $\varphi^\star g'=\Omega^2g$ and
$\xi'=\varphi_\star\xi$. Such a mapping projects to a
diffeomorphism of the quotients, $Q$ and $Q'$ we denote
by $\Phi$.
Then the same proof as in Ref. [5] allows one to show that
 if $(a'_\mu,\psi')$ is a solution of the field equations on $M'$, then
$$
a_\mu=(\varphi^\star a')_\mu
\and
\psi=\Omega\,\varphi^\star\psi'
\equation
$$
is a solution of the analogous equations on $M$. Locally we have
$\varphi(t,\vec{x},s)=(t',\vec{x}',s')$
with $(t',\vec{x}')=\Phi(t,\vec{x})$
and $s'=s+\Sigma(t,\vec{x})$
so that
$\psi=\Omega\,\varphi^\star\psi'$ reduces to $$
\Psi(t,\vec{x})=\Omega(t)\,e^{i\Sigma(t,\vec{x})}\Psi'
(t',\vec{x}'), \qquad
A_\alpha=\Phi^\star A'_\alpha
\qquad(\alpha=0,1,2).
\equation
$$
Note that $\varphi$ takes a $\xi$-preserving conformal
transformation of $(M,g,\xi)$ into a $\xi'$-preserving
conformal transformation of $(M',g',\xi')$.
Conformally related Bargmann spaces
have therefore isomorphic symmetry groups.

The associated conserved quantities can be related by
comparing the expressions in Eq.~(2.6). Note first that,
for $\psi$ as in Eq.~(2.10), $
D_\mu\psi=\Omega\,(\varphi^\star D'_\mu\psi')+ \Omega^{-1}
\nabla'_\nu\Omega\,\varphi^\star\psi' $.
Using
$
R=\Omega^2\,\varphi^\star R'+6\Omega^{-1}\nabla'_\nu\nabla'{}^\nu\Omega $
and
$
\xi_\mu=\Omega^{-2}\xi'_\mu
$
as well as $\Omega=\Omega(t)$ and
that $g^{\mu t}$ is non-vanishing only for $\mu=s$ one
finds hence that
$
\vartheta_{\mu\nu}\xi^\nu=\Omega^2\varphi^\star
\big(\vartheta'_{\mu\nu}\xi'{}^\nu\big). $
But
$
\sqrt{\gamma}=\Omega^{-2}\varphi^\star\sqrt{\gamma'} $.
Therefore, the conserved
quantity (2.5)
associated to $X=(X^\mu)$ on $(M,g,\xi)$ and to
$X'=\varphi_\star X$ on $(M',g',\xi')$ coincide,
$$
Q_X=\varphi^\star Q'_{X'}.
\equation
$$
The labels of the generators
are, however, different (see the examples below).

\goodbreak


\chapter{Flat examples}

Consider now the Lorentz metric
$$
d\vec{x}_{osc}^2+2dt_{\rm osc}ds_{\rm osc}-\omega^2r_{\rm osc}
^2dt_{\rm osc}^2 \equation
$$
where $\vec{x}_{\rm osc}\in\IR^2$, $r_{\rm osc}=
|\vec{x}_{\rm osc}|$ and $\omega$ is a constant.
Its null geodesics correspond to a non-relativistic,
spinless particle in an oscillator background [6,9].
Requiring equivariance (2.4), the wave equation (2.1) reduces to $$
i\partial_{t_{\rm osc}}\Psi_{\rm osc}=\left\{
-{\vec{D}^2\over2}+{\omega^2\over2}{r_{\rm osc}}^2
-\Lambda\,\Psi_{\rm osc}\Psi_{\rm osc}^*\right\}
\Psi_{\rm osc} \equation
$$
($\vec{D}=\vec{\partial}-i\vec{A}$, $\Lambda=\lambda/2$),
 which describes Chern-Simons vortices in a harmonic force
background, studied in Ref. [3].
The clue is that the mapping
$\varphi(t_{\rm osc},\vec{x}_{\rm osc},s_{\rm osc})
=(T,\vec{X},S)$ [9], where
$$
T={\tan\omega\,t_{\rm osc}\over\omega},
\qquad
\vec{X}={\vec{x}_{\rm osc}\over\cos\omega t_{\rm osc}}, \qquad
S=s_{\rm osc}-{\omega r_{\rm osc}^2\over2}\tan\omega t_{\rm osc}
 \equation
$$
carries the oscillator metric (3.1) conformally into the
free form, $
d\vec{X}^2+2dTdS
$,
with conformal factor
$\Omega(t_{\rm osc})=|\cos\omega t_{\rm osc}|^{-1}$ such
 that $\varphi_\star\partial_{s_{\rm osc}}=\partial_S$.
Our formula lifts the
coordinate transformation of Ref. [4] to Bargmann space.

A solution in the harmonic background can be obtained by Eq.~(2.11).
A subtlety arises, though:
the mapping (3.3) is many-to-one:
it maps each `open strip'
$$
I_j=\big\{
(\vec{x}_{\rm osc},t_{\rm osc},s_{\rm osc})\,\big| \,
(j-\2)\pi<\omega t_{\rm osc}<(j+\2)\pi \big\},
\equation
$$
where $\,j=0,\pm1,\ldots$ corresponding to a half
oscillator-period onto the full Minkowski space.
Application of (2.11) with $\Psi$ an `empty-space'
solution yields, in each $I_j$, a solution,
$\Psi^{(j)}_{\rm osc}$. However, at the contact points
$t_j\equiv(j+1/2)(\pi/\omega)$, these fields may not match.
 For example, for the `empty-space' solution obtained by
 an expansion, Eq.~(2.7) with $\vec{b}=0,\,k\neq0$,
$$
\lim_{t_{\rm osc}\to t_j-0}\Psi^{(j)}_{\rm osc}= (-1)^{j+1}
{\omega\over k}
e^{-i{\omega^2\over2k}r_{\rm osc}^2}\Psi_0(-{\omega\over k}
\vec{x})= -\lim_{t_{\rm osc}\to t_j+0}\Psi^{(j+1)}_{\rm osc}.
 \equation
$$
Then continuity is restored by including the `Maslov'
phase correction [10]:
$$
\left\{\eqalign{
\Psi_{\rm osc}(t_{\rm osc},\vec{x}_{\rm osc})&= (-1)^{j}\,
\displaystyle{1\over\cos\omega t_{\rm osc}}\,
\exp\left\{-{i\omega\over2}r_{\rm osc}^2\tan{\omega
t_{\rm osc}}\right\}\, \Psi(T,\vec{X})
\ccr
(A_{\rm osc})_0(t_{\rm osc},\vec{x}_{\rm osc})
&={1\over\cos^2\omega t_{\rm osc}}
\big[
A_0(T,\vec{X})-\omega\sin\omega t_{\rm osc}\;
\vec{x}_{\rm osc}\cdot\vec{A}(T,\vec{X}) \big],
\ccr
\vec{A}_{\rm osc}(t_{\rm osc},\vec{x}_{\rm osc})
&={1\over\cos\omega t_{\rm osc}}\,
\vec{A}(T,\vec{X}),
\cr}\right.
\equation
$$
where $j$ is as in (3.4).
Eq.~(3.6)
extends the result in Ref. [4] from $|t_{\rm osc}|<\pi/2\omega$
 to any $t_{\rm osc}$. For the static solution in [1] or for
that obtained from it by
a boost, $\lim_{t_{\rm osc}\to t_j}\Psi^{(j)}_{\rm osc}=0$
and the inclusion of the correction factor is not mandatory.

Chern-Simons theory in the oscillator-metric has again a
Schr\"odinger symmetry,
whose generators are related to those in `empty' space as
$$
\left\{
\eqalign{
J_{\rm osc}&=\cJ,
\cr
H_{\rm osc}&={\cH}+\omega^2\cK,
\cr
(C_{\rm osc})_\pm&=
\left(\cH-\omega^2\cK\pm 2i\omega\cD\right), \cr
(\vec{P}_{\rm osc})_\pm&=
\left(\vec{\cP}\pm i\omega\vec{\cB}\right), \cr
N_{\rm osc}&=\cM.
\cr
}\right.
\equation
$$
The oscillator-Hamiltonian, $H_{\rm osc}$, is hence a
combination of the Hamiltonian and of the expansion
valid for $\omega=0$, etc. The generators $H_{\rm osc}$
 and $(C_{\rm osc})_\pm$ span ${\rm o}(2,1)$ and the
$(\vec{P}_{\rm osc})_\pm$ generate the two-dimensional
Heisenberg algebra [9].
Eq.~(3.7) adds $(C_{\rm osc})_\pm$ and $(\vec{P}_{\rm osc}
)_\pm$ to the $J_{\rm osc}$ and $H_{\rm osc}$ in Ref.~[3].

Consider next the metric
$$
d\vec{x}{}^2+2dt\Big[ds+
\2\epsilon_{ij}{\cal B}{x}^jd{x}^i\Big]
\equation
$$
where $\vec{x}\in\IR^2$ and ${\cal B}$
is a constant. Its null geodesics
describe a charged particle in a uniform magnetic field
in the plane [6].
Again, when
imposing equivariance, Eq.~(2.1)
reduces precisely
to Eq.~(1.1) with $\Lambda=\lambda/2$ and covariant
derivative $D_\omega$ given as in Eq.~(1.2). The metric
 (3.8) is readily transformed into an oscillator metric (3.1):
the mapping
$\varphi(t,\vec{x},s)=(t_{\rm osc},\vec{x}_{\rm osc},
s_{\rm osc})$ given by
$$
t_{\rm osc}=t,
\qquad
x_{\rm osc}^i=x^i\cos\omega t+\epsilon^i_jx^j\sin\omega t,
 \qquad
s_{\rm osc}=s
\equation
$$
--- which amounts to switching
to a rotating frame with angular velocity $\omega={\cal B}/2$
 --- takes the \lq constant ${\cal B}$-metric' (3.8) into the
 oscillator metric (3.1).
The vertical vectors
$\partial_{s_{\rm osc}}$ and $\partial_s$ are permuted. Thus,
 the time-dependent rotation (3.9)
followed by the transformation (3.3),
which projects to the coordinate
transformation (1.5) of Refs. [2] and [3], carries conformally the
constant-${\cal B}$ metric (3.8) into the $\omega\=0$-metric. It
carries therefore
the \lq empty' space so\-lution $e^{is}\Psi$ with $\Psi$
 as in (2.7) into that in
a uni\-form mag\-netic field back\-ground according to
Eq.~(2.10). Taking into account the equivariance, we get
 the formul{\ae} of [2] i.e. (1.4) without the
${\cal N}$-terms --- but multiplied with the Maslov
factor $(-1)^j$. (The
${\cal N}$-term arises due to a subsequent gauge
transformation required by the gauge fixing in [3]).

It also allows to
\lq export' the Schr\"odinger symmetry to non-relativistic
 Chern-Simons theory in
the constant magnetic field background. The [rather
complicated] generators, listed in Ref. [11], are readily
obtained using Eq.~(2.12). For example, time-translation
$t\to t+\tau$ in the ${\cal B}$-background amounts to a
time translation for the oscillator with parameter $\tau$
plus a rotation with angle $\omega\tau$. Hence
$
H_{\cal B}=H_{\rm osc}-\omega\cJ=\cH+\omega^2\cK-\omega\cJ. $
Similarly, a space translation for ${\cal B}$ amounts,
in `empty' space, to a space translations and a boost,
followed by a rotation, yielding $P_B^i=\cP^i+\omega\,
\epsilon^{ij}\cG^j$, etc.

\goodbreak


\chapter{Conformally flat Bargmann spaces}

All our preceding results apply to any Bargmann space
which can be conformally mapped into Minkowski space in
 a $\xi$-preserving way. Now we describe these
`Schr\"odinger-conformally flat' spaces. In $D=n+2>3$
dimensions, conformal
flatness is guaranteed by the
vanishing of the conformal Weyl tensor
$$
C^{\mu\nu}_{\ \ \rho\sigma}
=
R^{\mu\nu}_{\ \ \rho\sigma}
-
\smallover 4/{D-2}\,\delta^{[\mu}_{\ [\rho}\,R^{\nu]}_{\ \sigma]} +
\smallover 2/{(D-1)(D-2)}\,
\delta^{[\mu}_{\ [\rho}\,\delta^{\nu]}_{\ \sigma]}\,R. \equation
$$
Now $R_{\mu\nu\rho\sigma}\xi^\mu\equiv 0$ for a Bargmann
 space, which implies some extra conditions
on the curvature. Inserting the identity $\xi_\mu
R^{\mu\nu}_{\ \ \rho\sigma}=0$
into
$C^{\mu\nu}_{\ \ \rho\sigma}=0$,
using the identity $\xi_\mu R^\mu_\nu\equiv 0$
($R^\nu_\sigma\equiv R^{\mu\nu}_{\ \ \mu\sigma}$), we find
$
0=
-\left[
\xi_\rho R^\nu_\sigma-\xi_\sigma R^\nu_\rho\right] +
R/(D-1)\left[
\xi_\rho\delta^\nu_\sigma-\xi_\sigma\delta^\nu_\rho \right]
$.
Contracting again with $\xi^\sigma$ and using that
$\xi$ is null, we end up with
$R\,\xi_\rho\xi^\nu=0$. Hence the scalar curvature
vanishes, $R=0$. Then the previous equation yields
$\xi^{ }_{[\rho}R^\nu_{\sigma]}=0$ and thus
$R_\sigma^\nu=\xi_\sigma\eta^\nu$
for some vector field $\eta$. Using the
symmetry of the Ricci tensor, $R_{[\mu\nu]}=0$, we
 find that $\eta=\varrho\,\xi$
for some function $\varrho$. We finally get the
{\it consistency relation} $$
R_{\mu\nu}=\varrho\,\xi_\mu\xi_\nu.
\equation
$$
The Bianchi identities ($\nabla_\mu R^\mu_\nu=0$
since $R=0$) yield $\xi^\mu\partial_\mu\varrho=0$,
i.e. $\varrho$ is a function on spacetime $Q$.
The conformal Schr\"odinger-Weyl tensor is hence of the form
$$
C^{\mu\nu}_{\ \ \rho\sigma}=
R^{\mu\nu}_{\ \ \rho\sigma}
-
\smallover 4/{D-2}\,
\varrho\,\delta^{[\mu}_{\ [\rho}\,\xi^{\nu]}\xi^{ }_{\sigma]}.
 \equation
$$
It is noteworthy that Eq.~(4.2) is
the Newton-Cartan field equation with
$\varrho/(4\pi G)$ as the matter density of the sources.

It follows from Eq.~(4.2) that the transverse Ricci tensor
of a Schr\"odinger-conformal flat Bargmann metric necessarily
 vanishes, $R_{ij}=0$ for each~$t$.
The transverse space is hence (locally) flat and we can
choose $g_{ij}=g_{ij}(t)$.
Then a change of coordinates
$(t,\vec{x},s)\to(t,G(t)^{1/2}\vec{x},s)$ where $G(t)^i_j=
g_{ij}(t)$ [which brings in a uniform magnetic field and/or
 an oscillator to the metric],
casts our Bargmann metric into the form (2.8) with $g_{ij}
=\delta_{ij}$ while $\xi$ remains unchanged.

The non-zero components of the Weyl tensor of the general
$D=4$ Brinkmann metric (2.8) are found as $$
\eqalign{
&C_{xyxt}=-C_{ytts}=-\smallover1/4\partial_x{\cal B}, \ccr
&C_{xyyt}=+C_{xtts}=-\smallover1/4\partial_y{\cal B}, \ccr
&C_{xtxt}=-\2\big[
\partial_t(\partial_y{\cal A}_y-\partial_x{\cal A}_x)-
{\cal A}_x\partial_y{\cal B},
\big]
+\2\big[\partial_x^2-\partial_y^2\big]U, \ccr
&C_{ytyt}=+\2\big[
\partial_t(\partial_y{\cal A}_y-\partial_x{\cal A}_x)
-{\cal A}_y\partial_x{\cal B}\big]
-\2\big[\partial_x^2-\partial_y^2\big]U,\ccr &C_{xtyt}=+\2\big[
\partial_t(\partial_x{\cal A}_y+\partial_y{\cal A}_x)
+2\partial_x\partial_yU\big]
-\smallover1/4({\cal A}_x\partial_x-{\cal A}_y\partial_y){\cal B}.
 \cr
}
\equation
$$
Then Schr\"odinger-conformal flatness requires
$$
\left\{\eqalign{
&{\cal A}_i=\2\epsilon_{ij}{\cal B}(t)x^j+a_i, \qquad
\vec{\nabla}\times\vec{a}=0,
\qquad\partial_t\vec{a}=0,
\ccr
&U(t,\vec{x})=\2 C(t)r^2+\vec{F}(t)\cdot\vec{x}+K(t). \cr} \right.
\equation
$$
(Note, {\sl en passant}, that (4.2) automatically holds:
the only non-vanishing component of the Ricci tensor is $R_{tt}=
-\partial_t(\vec{\nabla}\cdot\vec{\cal A})-\2{\cal B}^2-\Delta U$).

This Schr\"odinger-conformally flat metric hence allows
one to describe a uniform magnetic field ${\cal B}(t)$,
an attractive [$C(t)=\omega^2(t)$] or repulsive [$C(t)=
-\omega^2(t)$]
isotropic oscillator and a uniform force field $\vec{F}
(t)$ in the plane which may all depend arbitrarily on time.
 It also includes a curlfree vector potential~$\vec{a}
(\vec{x})$ that can be gauged away if the transverse space
 is simply connected: $a_i=\partial_if$ and the coordinate
 transformation $(t,\vec{x},s)\to(t,\vec{x},s+f)$ results
in the `gauge'
transformation ${\cal{A}}_i\to{\cal{A}}_i-\partial_if=
-\2{\cal B}\,\epsilon_{ij}x^j$. However, if space is
not simply connected, we can also include an external
Aharonov-Bohm-type vector potential.

Being conformally related, all these metrics share the
symmetries of flat Bargmann space: for example, if the
transverse space is $\IR^2$ we get the full Schr\"odinger
symmetry; for $\IR^2\setminus\{0\}$ the symmetry is
reduced rather to
${\rm o}(2)\times{\rm o}(2,1)\times\IR$, just like
for a magnetic vortex [12].

The case of a constant electric field
is quite amusing. Its metric,
$d\vec{x}^2+2dtds-2\vec{F}\cdot\vec{x}dt^2$, can be
brought to the free form by switching to an accelerated
coordinate system, $$
\vec{X}=\vec{x}+\2\vec{F}\,t^2,
\qquad
T=t,
\qquad
S=s-\vec{F}\cdot\vec{x}\,t-\smallover1/6\vec{F}^2t^3.
\equation
$$
This example (chosen by Einstein to illustrate the equivalence
 principle) also shows that the action of the Schr\"odinger
 group --- e.g. a rotation --- looks quite different in the
 inertial and in the moving frames.

Let us finally mention that the above results admit a `gauge
 theoretic' interpretation. In conformal (Lorentz) geometry,
 the Weyl tensor $C_{\mu\nu\rho\sigma}$ arises as part of the
 ${\rm o}(n+2,2)$-valued curvature
of a Cartan connection for a $D=n+2$ dimensional base manifold.
 The Schr\"odinger-conformal geometry in this dimension can
be viewed as a reduction of the standard conformal geometry
to the Schr\"odinger subgroup ${\rm Sch}(n+1,1)\subset{\rm O}
(n+2,2)$
[13]. The curvature of the reduced Cartan connection then
defines the Schr\"odinger-Weyl tensor which is thus characterized by
$$
C_{\mu\nu\rho\sigma}\,\xi^\mu=0,
\equation
$$
a property coming from the previous embedding and strictly
 equivalent to Eqs.~(4.2,3).

\chapter{Conclusion}

Our `non-relativistic Kaluza-Klein' approach provides a
unified view on the various vortex constructions in external
 fields, explains the common origin of the large symmetries,
and allows us to describe all such spaces. We have also
pointed out that the formula (1.4) may require a slight
modification for times larger then a half oscillator-period.

\kikezd{Acknowledgement}. We are indebted to Professor
R. Jackiw for stimulating discussions.
L. P. would like to thank Tours University for the hospitality
 extended to him
and to the Hungarian National Science and Research Foundation
 (Grant No. $2177$) for a partial financial support.

\vfill\eject
\goodbreak
\vskip5mm


\reference
R. Jackiw and S-Y. Pi,
Phys. Rev. Lett. {\bf 64}, 2969 (1990);
Phys. Rev. {\bf 42}, 3500 (1990).

\reference
Z. F. Ezawa, M. Hotta and A. Iwazaki,
Phys. Rev. Lett. {\bf 67}, 411 (1991);
Phys. Rev. {\bf D44}, 452 (1991).

\reference
R. Jackiw and S-Y. Pi,
Phys. Rev. Lett. {\bf 67}, 415 (1991).

\reference
R. Jackiw and S-Y. Pi, Phys. Rev. {\bf 44}, 2524 (1991).

\reference
C. Duval, P. A. Horv\'athy and L. Palla, Phys. Lett.
{\bf B325}, 39 (1994).

\reference
C. Duval, G. Burdet, H-P. K\"unzle and M. Perrin, Phys.
Rev. {\bf D31}, 1841 (1985); see also C. Duval, G. Gibbons
 and P. Horv\'athy,
Phys. Rev. {\bf D43}, 3907 (1991).

\reference
H.W. Brinkmann, Math. Ann. {\bf 94}, 119 (1925).

\reference
R. Jackiw, Phys. Today {\bf 25}, 23 (1972); U. Niederer,
Helv. Phys. Acta {\bf 45}, 802 (1972); C. R. Hagen, Phys.
 Rev. {\bf D5}, 377 (1972); G. Burdet and M. Perrin, Lett.
 Nuovo Cim. {\bf 4}, 651 (1972).

\reference
U. Niederer, Helv. Phys. Acta {\bf 46}, 192 (1973);
G. Burdet, C. Duval and M. Perrin,
Lett. Math. Phys. {\bf 10}, 255 (1986);
J. Beckers, D. Dehin and V. Hussin, J. Phys. {\bf A20},
 1137 (1987).

\reference
V. P. Maslov, {\it The Theory of Perturbations and
Asymptotic Methods} (in Russian), izd. MGU: Moscow, (1965);
J-M Souriau, in Proc {\it Group Theoretical Methods
in Physics}, Nijmegen'75, Janner (ed) Springer Lecture
Notes in Physics {\bf 50}, (1976).

\reference
M. Hotta, Prog. Theor. Phys. {\bf 86}, 1289 (1991).

\reference
R. Jackiw, Ann. Phys. (N. Y.) {\bf 201}, 83 (1990).

\reference
M. Perrin, G. Burdet and C. Duval,
Class. Quant. Grav. {\bf 3}, 461 (1986); C. Duval, in
{\it Conformal Groups and Related Symmetries. Physical
Results and Mathematical Background}, ed. A. O. Barut
and H. D. Doebner, Lecture Notes in Physics {\bf 261},
p. 162 Berlin: Springer-Verlag (1986).

\vfill\eject

\bye